\documentclass[seceq]{ptptex}

\usepackage{graphicx}

\newcommand{\be}{\begin{equation}}
\newcommand{\ee}{\end{equation}}
\newcommand{\ba}{\begin{eqnarray}}
\newcommand{\ea}{\end{eqnarray}}

\newcommand{\rth}{\frac{1}{\sqrt{3}}}
\newcommand{\rsix}{\frac{1}{\sqrt{6}}}

\newcommand{\pr}{^\prime}

\newcommand{\Sg}{\Sigma^*}

\newcommand{\X}{\Xi^*}

\title{Recent developments in chiral dynamics of
hadrons and hadrons in nuclei
}

\author{ \scshape  
E. \textsc{Oset}$^{1}$,\footnote{ e-mail address:
oset@ific.uv.es}  
M. \textsc{Doring}$^{1}$, M. \textsc{Kaskulov}$^{1}$, L. \textsc{Roca}$^{1}$,
S. \textsc{Sarkar}$^{1}$, D. \textsc{Strottman}$^{1}$, M.J.
\textsc{Vicente Vacas}$^{1}$,
V.K. \textsc{Magas}$^{2}$, A. \textsc{Ramos}$^{2}$, 
and E. \textsc{Hernandez}$^{3}$
}

\inst{
$^1$  Departamento de F\'{\i}sica Te\'orica and IFIC, Centro Mixto,\\ 
     Institutos de Investigaci\'on de Paterna - Universidad de Valencia-CSIC\\ 
    $^2$ Departament d'Estructura i 
Constituents de la Mat\'eria, 
Universitat de Barcelona,
 Diagonal 647, 08028 Barcelona, Spain\\
 $^3$ Grupo de F\'{\i}sica Nuclear, Departamento de Fisica Fundamental
e IUFFyM, \\
\small Facultad de Ciencias, 
Universidad de Salamanca, 
}

\abst{In this talk I present recent developments in the field of hadronic
physics and hadrons in the nuclear medium. I review the  unitary chiral 
approach to meson baryon interaction and address the topics of the 
 two dynamically generated $\Lambda(1405)$ resonances, with  experiments 
 testing it, the $\Lambda(1520)$ and $\Delta(1700)$ resonances, plus the 
$\Lambda(1520)$, $\Sigma(1385)$ and $\omega$ in the nuclear medium.
}

\begin{document}

\maketitle

\section{Introduction}

The unitary extrapolations of chiral perturbation theory are having a great
impact in hadronic and nuclear physics.  They allow one to address the meson
meson  and meson baryon interactions at low and intermediate energies and have
also shown that many known mesonic and baryonic resonances naturally appear as
a consequence of these interactions, hence producing dynamically generated
states which go beyond the usual $q \bar{q}$ or $qqq$ nature of the mesons and
baryons.  The identification of such resonances and the study of their
properties and their influence in different reactions is a growing field
allowing to interpret many phenomena and  making
predictions on observables of the resonances and special
features in  reaction dynamics. The main ideas behind this chiral unitary 
theory are reviewed in \cite{review}. A more recent review on applications 
to physical processes can also be found in \cite{puri}.

\section{Chiral unitary approach}

   Although much work has been done along these lines, we follow here the basic
ideas of the N/D method, adapted for the meson meson interaction  using 
chiral dynamics in \cite{nsd}, and to the meson baryon case in \cite{joseulf}.  
 A pedagogical description of the method and applications to physical cases can
 be seen in \cite{puri}. The essence of the method consists in the use of the
 chiral Lagrangians as a source of dynamical interaction, then implement
 unitarity in coupled channels to get $Im T^{-1}$ from the phase space of the
 intermediate states of the coupled channels and then implement a dispersion
 relation to get the real part, to which $V^{-1}$ is added, with $V$ the tree 
 level amplitude obtained from the chiral Lagrangians. The method requires a
 subtraction constant in the dispersion relation and one has an idea of its
 order of magnitude for it to have a natural size.
   Applications to the interaction of $\bar{K} N$  and coupled channels  
   is discussed
in \cite{puri}, together with the  $K^-$-nuclear system and other systems and
reactions. We take advantage here to present a qualitative  description of the
method followed to study the interaction of the octet of pesudoscalar mesons with
the decuplet of baryons, as done in \cite{kolo} and \cite{sarkar}.
 
      The lowest order term of the chiral Lagrangian relevant for the interaction 
      of the baryon decuplet with the octet of pseudoscalar mesons is given 
by~\cite{Jenkins:1991es}.
\be
{\cal L}=-i\bar T^\mu {\cal D}\!\!\!\!/ T_\mu 
\label{lag1} 
\ee
where $T^\mu_{abc}$ is the spin decuplet field and $D^{\nu}$ the covariant derivative
given in \cite{sarkar}
The formalism provides the identification of the $SU(3)$ component
of $T$ to the physical states:
$T^{111}=\Delta^{++}$, $T^{112}=\rth\Delta^{+}$, $T^{122}=\rth\Delta^{0}$,
$T^{222}=\Delta^{-}$, $T^{113}=\rth\Sigma^{*+}$, $T^{123}=\rsix\Sigma^{*0}$,
$T^{223}=\rth\Sigma^{*-}$,  $T^{133}=\rth\Xi^{*0}$,
$T^{233}=\rth\Xi^{*-}$, $T^{333}=\Omega^{-}$.

From this Lagrangian, for a meson of incoming (outgoing) momenta $k(k\pr)$ one obtains  the 
simple form for the $S$-wave transition amplitudes
\be
V_{ij}=-\frac{1}{4f^2}C_{ij}(k^0+k^{\pr 0}).
\label{poten}
\ee 

The coefficients $C_{ij}$ for reactions with all possible values of
strangeness $(S)$ and charge $(Q)$ are given in Appendix-I and II of \cite{sarkar}. 
We then consider
$S=-1$ in $I=0$. In this case there are the $\Sg\pi$ and $\X K$ states in the
coupled channels states.
 The coefficients are given in \cite{sarkar} and the interaction is found
 attractive, leading to a resonance which can be identified with the
 $\Lambda(1520)$ and would be essentially a bound state of the $\Sg\pi$.

The results of the $|T|^2$ are shown in \cite{sarkar}. There one can see
peaks which can be associated to different resonances. Here in this talk I will
only discuss the $\Lambda(1520)$ and the $\Delta(1700)$.

\section{The two $\Lambda(1405)$ states dynamically generated} 

We come back to the interaction of the octet of pseudoscalar mesons with the
octet of baryons and refer to the work \cite{jido} where two $\Lambda(1405)$
states were found. In an SU(3) symmetric world the combination of $8\times 8$
representation gives rise to the irreducible representations:
$1$, $8_s$, $8_a$, $10$, $\overline{10}$ and $27$. The chiral Lagrangians
(lowest order chiral Lagrangian) that
we use are SU(3) symmetric and only the different masses of the hadrons break
this symmetry after unitarization. 
 If we do an SU(3) symmetric approximation, making the masses of the baryons
 equal and the masses of the mesons equal, 
and look for poles of the scattering matrix, we find poles
corresponding to the octets, which are degenerate in this limit, and the singlet.

\begin{figure}
\centering
\includegraphics[width=8.0cm]{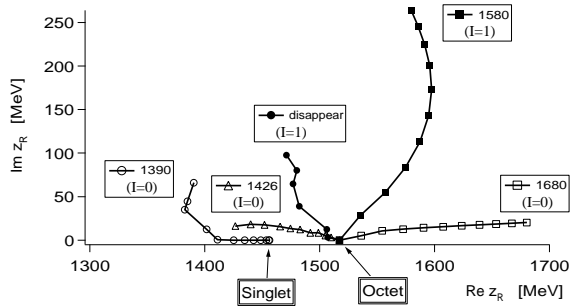}
  \caption{Trajectories of the poles in the scattering amplitudes obtained by
  changing the SU(3) breaking parameter $x$ gradually. At the SU(3) symmetric 
  limit ($x=0$),
   only two poles appear, one is for the singlet and the other for the octet.
  The symbols correspond to the step size $\delta x =0.1$. }
  \label{fig:tracepole}
\end{figure}

By  using the
physical masses of the baryons and the mesons, the position of the
poles change and the two octets split apart in four branches, two
for $I=0$ and two for $I=1$, as one can see in \cite{jido},
 which we reproduce in 
Fig.~\ref{fig:tracepole}. In the figure we show the trajectories
of the poles as a function of a parameter $x$ that breaks
gradually the SU(3) symmetry up to the physical values.  

The splitting of the two $I=0$ octet states is very interesting.
One moves to higher energies to merge with the $\Lambda(1670)$
resonance and the other one moves to lower energies to create a
pole, quite well identified below the  $\bar{K}N$ threshold, with
a narrow width. 
 On the other hand, the singlet also evolves
to produce a pole at low energies with a quite large width.

We note that the singlet and the $I=0$ octet states appear
nearby in energy and  what experiments  actually
see is a combination of the effect of these two resonances. We should also note
that the narrow state at higher energy couples mostly to $\bar{K} N$ and the wide
one at lower energies couples mostly to $\pi \Sigma$. Recent works 
\cite{borasoyw,ollerpra,ollersolo,borasoyulf} include contributions from higher order
Lagrangians and they also have the
narrow pole very similar to \cite{jido},  the wide pole also appearing
at lower energies than the narrow one, but there is some dispersion in the 
actual values of the width. The results for the amplitudes obtained with the
lowest order Lagrangian \cite{jido} fall within the theoretical uncertainties of
the more complete models \cite{borasoyulf}.

\begin{figure}[htb]
\centering
\includegraphics[width=4.0cm]{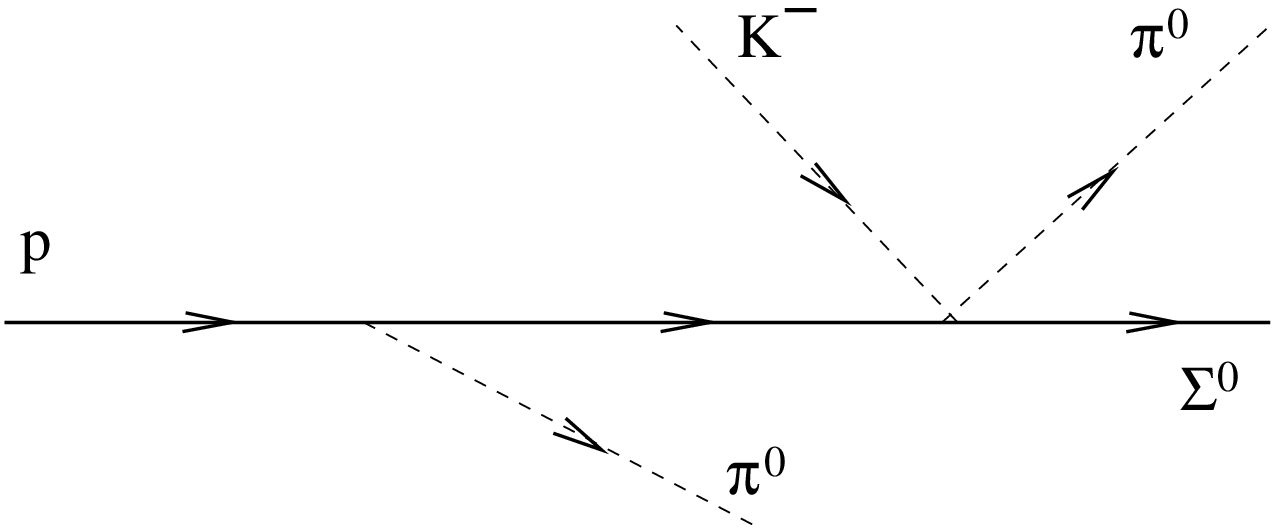}
\hspace{0.2cm}
\includegraphics[width=5.0cm]{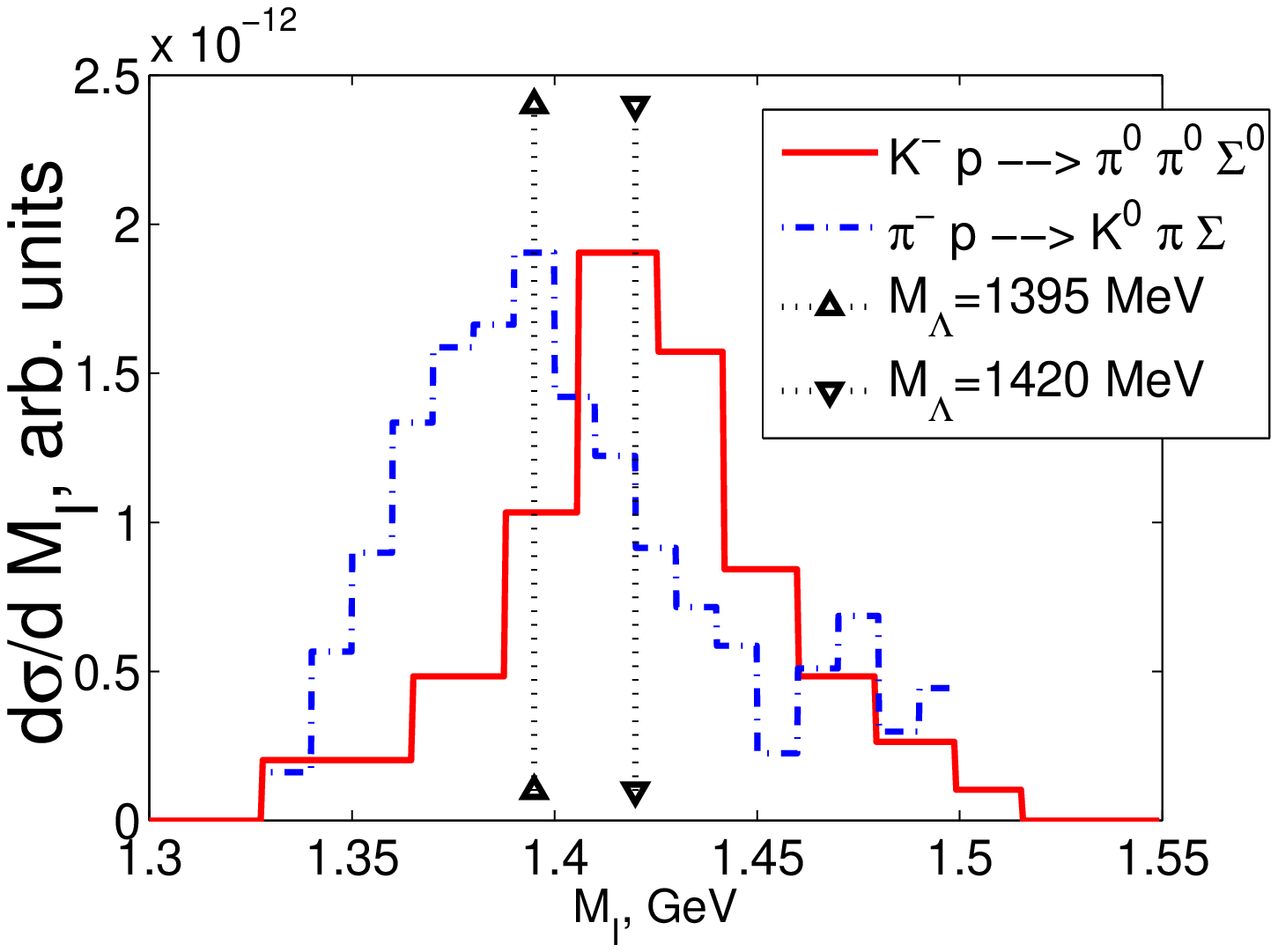}
 \caption{Left: Nucleon pole term for the $K^- p \to \pi^0 \pi^0 \Sigma$ reaction. 
 Right: Two experimental shapes of  $\Lambda(1405)$ resonance. 
 See text for more details. 
 \label{two_exp}
 }
\vspace{-0.5cm}
\end{figure}

  The recently measured reaction
$K^- p \to \pi^0 \pi^0 \Sigma^0$  \cite{Prakhov} allows us to test already the
two-pole nature of the $\Lambda(1405)$. This
process
shows a strong similarity with the reaction $K^- p \to \gamma \Lambda(1405)$, where the
photon is replaced by a $\pi^0$, and which was studied in \cite{nacher}.
Our model \cite{magas} for the reaction 
$K^- p \to \pi^0 \pi^0 \Sigma^0 $
in the energy region of $p_{K^-}=514$ to $750$ MeV/c, as in the experiment \cite{Prakhov}, 
considers those mechanisms in which a $\pi^0$ loses the necessary energy
to allow the remaining $\pi^0\Sigma^0$ pair to be on top of the $\Lambda(1405)$
resonance.  The most important mechanisms is given by the diagram of
Fig.~\ref{two_exp}.
  As a consequence, the $\Lambda(1405)$ thus obtained comes
mainly from the $K^- p \to \pi^0 \Sigma^0$ amplitude which, as mentioned above,
gives  the largest possible weight  to the second (narrower) state.


 We recall
that the shape of the  $\Lambda(1405)$ in the $\pi^- p \to K^0 \pi \Sigma$ 
reaction was shown in Ref.~\cite{hyodo} to be largely built from the
 $\pi \Sigma \to \pi \Sigma$ amplitude, which is dominated by
the wider, lower energy state.  In Fig.~ \ref{two_exp} we show the experimental
results for both experiments, showing a quite different shape in consonance with
the theoretical findings of \cite{hyodo} and \cite{magas}.
The quite different shapes of the 
 $\Lambda(1405)$ resonance seen in these experiments can be interpreted in favour
 of the existence of two poles with the corresponding states having the
 characteristics predicted by the chiral theoretical calculations.

 It is interesting to see that a similar shape for the $\Lambda(1405)$ was seen
 some time ago in the reaction $K^- p \to \gamma \pi \Sigma$ \cite{nacher}.
   In this respect it would be interesting to carry our this reaction
   experimentally.  We also have here the suggestion for further reactions which
   could provide extra evidence on this issue. These would be the reactions:

 $K^- p \to \pi^0 \gamma \Lambda$ and  $\pi^- p \to  K^0 \gamma \Lambda$ 
 ~~looking for the $(\gamma \Lambda )$ invariant mass, plus the 
 
 $K^- p \to \pi^0 \gamma \Sigma^0$  and $\pi^- p \to  K^0 \gamma \Sigma^0$ 
 ~~looking for the $(\gamma \Sigma)$ invariant mass.

   This would provide the radiative decay of the  $\Lambda(1405)$ into 
   $\gamma \Lambda$ and  $\gamma \Sigma^0$ with the likely result of finding 
different rates depending on the reaction because of the two  
   $\Lambda(1405)$ states.

\section{The $\Lambda(1520)$}

\begin{figure}[h]
\centering
\includegraphics[width=3.5cm]{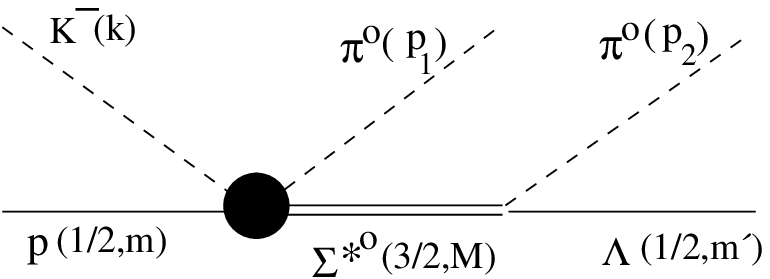}
\hspace{0.2cm}
\includegraphics[width=6.0cm]{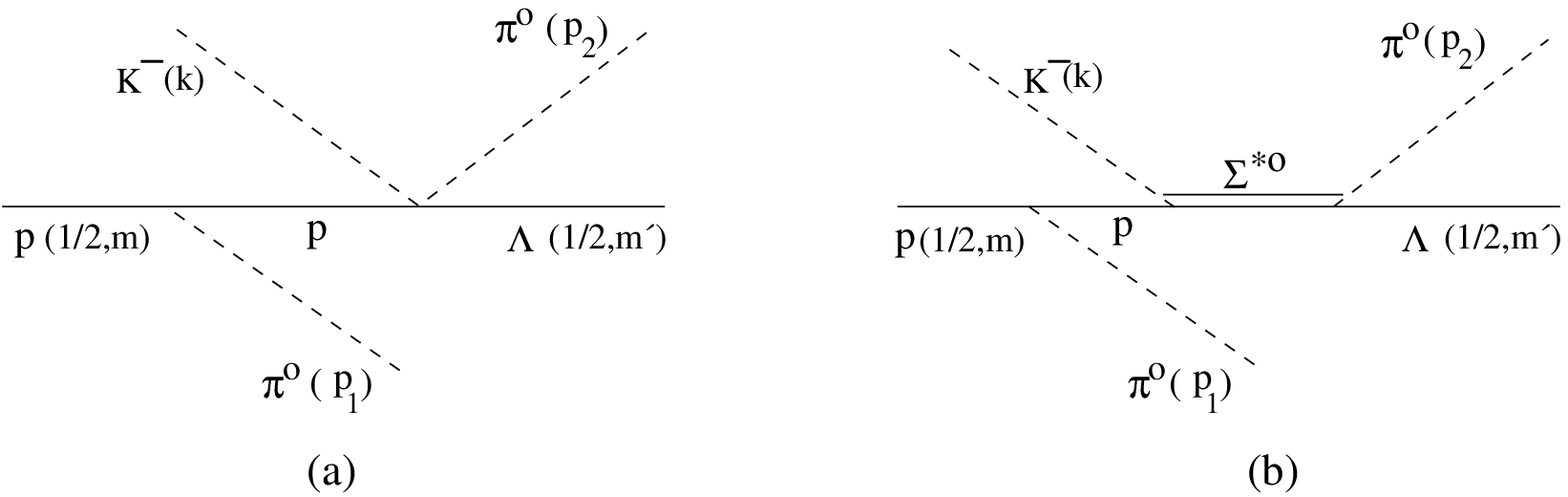}
\caption{Scheme for $K^-p \to \pi^0\Sigma^{*0}(1385) \to \pi^0\pi^0\Lambda$. The
blob indicates the unitarized vertex. Last two diagrams for conventional
background}
\label{kpfig}
\end{figure}

\begin{figure}[h]
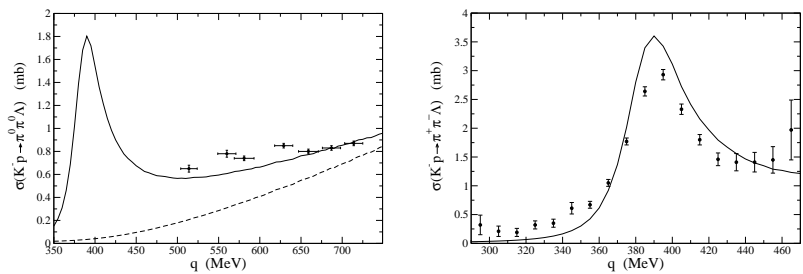

\centering
\includegraphics[width=5.0cm]{fig66.eps}
\hspace{0.3cm}
\includegraphics[width=5.0cm]{fig7.eps}
\caption{Result for the $K^-p\to\pi^0\pi^0\Lambda$ cross section.
Experimental data from Ref.~\cite{Prakhov:2004ri}}
\label{fig:nefkens}
\end{figure}

As we have seen in section 2, the channels $\pi \Sigma(1385)$ and $K \Xi$ form
the basis for the $S=-1$ and $I=0$.  The interaction in this sector is
attractive  and a resonance pole appears which can be brought close
to the $\Lambda(1520)$ mass with
fine tuning of the subtraction constants in the dispersion relation.  Yet, with
these channels and the position of the resonance at the experimental mass, the
resonance would not have a width and would qualify as a bound state of the 
$\pi \Sigma(1385)$.  This is of course rough and one should include the
$\bar{K}N$ and $ \pi \Sigma$ channels which appear in D-wave and which provide
the main source of decay of the $\Lambda(1520)$. The inclusion of the D-wave
channels is done in \cite{souravd,luisd}. There the couplings provided by the
chiral Lagrangian between the  $\pi \Sigma(1385)$ and $K \Xi$ states is kept and
the couplings to the D-wave states are fitted to experimental amplitudes. The
complete model shows that the coupling of the $\Lambda(1520)$ to the 
$\pi \Sigma(1385)$ state is still the largest, and thus the realistic world still
keeps some memory of the simplified one with only the s-wave channels. The
complete models allow us to address several reactions discussed in 
\cite{souravd,luisd}. 
In Fig.~\ref{fig:nefkens} 
 we show our results for 
$K^-p\to\pi^0\pi^0\Lambda$ and $K^-p\to\pi^+\pi^-\Lambda$
 cross section respectively along with experimental data
from Refs.~\cite{Prakhov:2004ri,Mast:1973gb}.
The dashed line represents the contribution from mechanisms 
 other than the unitarized
coupled channels, (a) and (b) of Fig.~\ref{kpfig} and the solid one gives the 
coherent sum of all
the processes, which includes the  amplitude of the left in Fig.~\ref{kpfig},
where the chiral dynamics of the $\bar{K} N \to \pi \Sigma^*$ transition appears.
Note that the cross section of the $K^-p\to\pi^+\pi^-\Lambda$ reaction
is a factor two larger at the peak than the $K^-p\to\pi^0\pi^0\Lambda$
one.
Hence, the agreement with the data is  a non-trivial accomplishment 
of the theory  since the $\bar{K} N \to \pi \Sigma^*$
amplitude has not been included in the fit. It would be interesting to measure
the  reaction of \cite{Prakhov:2004ri} at lower energies to get the
$\Lambda(1520)$ peak, something possible at JPARC.

Other issues related to the $\Lambda(1520)$ which have also been addressed within
the chiral unitary approach are the coupling of the $K^*$ to the resonance and
the radiative decay.  The first topic is addressed in \cite{hosaka} and the
coupling is compared to that obtained from ordinary quark models.
 The second topic is addressed in 
\cite{michasourav}, where one finds that the radiative decay of the
$\Lambda(1520)$ to $\Sigma \gamma$ is well reproduced, but the one to 
$\Sigma \gamma$ is quite short of the experimental results.  This magnitude
 becomes a good test of
possible extra components of the $\Lambda(1520)$, like some 3q components. 
Comparison of the results with quark model results \cite{bernard}, and some 
evaluations done in \cite{michasourav}, lets one conclude that a small fraction of
a 3q component, packed like in the chiral quark models, in addition to the meson
cloud component of the dynamically generated state, could bring the results in
agreement with data, but more work is needed there. 

\section{The $\Delta(1700)$ resonance}
The $\Delta(1700)$ is another of the resonances coming from the interaction of
the octet of mesons with the baryon decuplet. 
The results for the different channels, as studied in \cite{pieta} and
\cite{delta} are by no means trivial. Should we assume that the $\Delta(1700)$
belongs to an SU(3) decuplet, as suggested in the PDG, it is easy to see, using
SU(3) Clebsch Gordan coefficients, that
the couplings to the $\Delta \pi$, $\Sigma^* K$ and $\Delta \eta$ states in
$I=3/2$ would be proportional to 1, 2/5, 1/5 . However, with the dynamically
generated $\Delta(1700)$ they are proportional to 1,11.56 and 4.84 respectively.
Hence, one assumption or the other lead to differences in the cross sections of
the order of a factor 400-900 depending on the channel.  This said, the
agreement found in \cite{delta}, within theoretical and
experimental uncertainties, with the data for photon and pion induced
reactions having the $\Delta(1700)$ in the entrance
channel, must be considered a real success.  Below we plot
some results. In addition there are recent 
preliminary data for
these reactions at ELSA, with smaller errors, from  F. Klein et al. for the
reactions of the first figure, and from M. Navova et al.  for the other two
reactions which fall within the shaded region of our
calculations. The theoretical band comes from uncertainties in input from the PDG
used, which could be reduced in the future with improvements in the measurements
of some partial decay widths.

\begin{figure}[h]
\includegraphics[width=4.0cm]{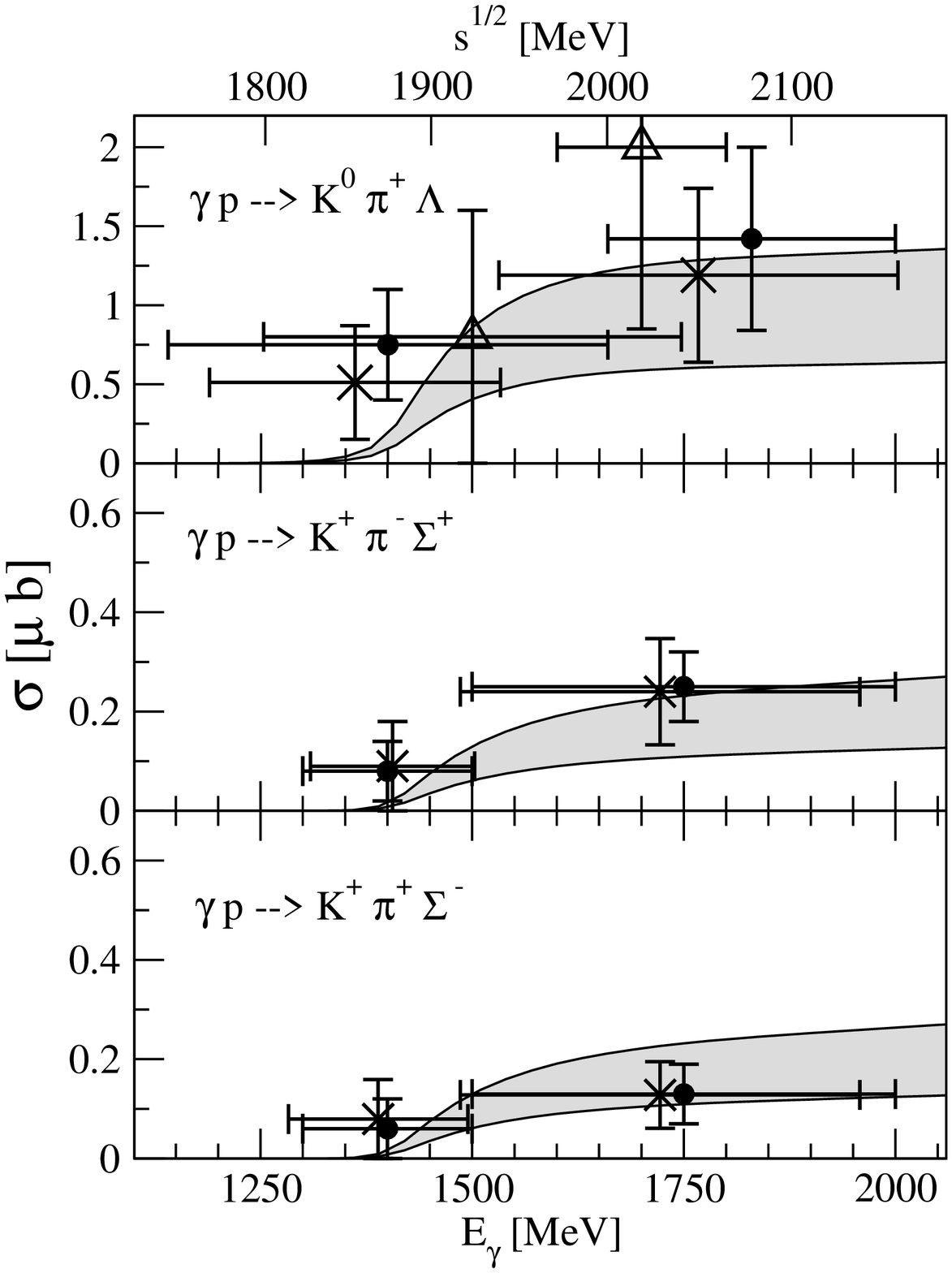}
\hspace*{0.2cm}
\includegraphics[width=4.2cm]{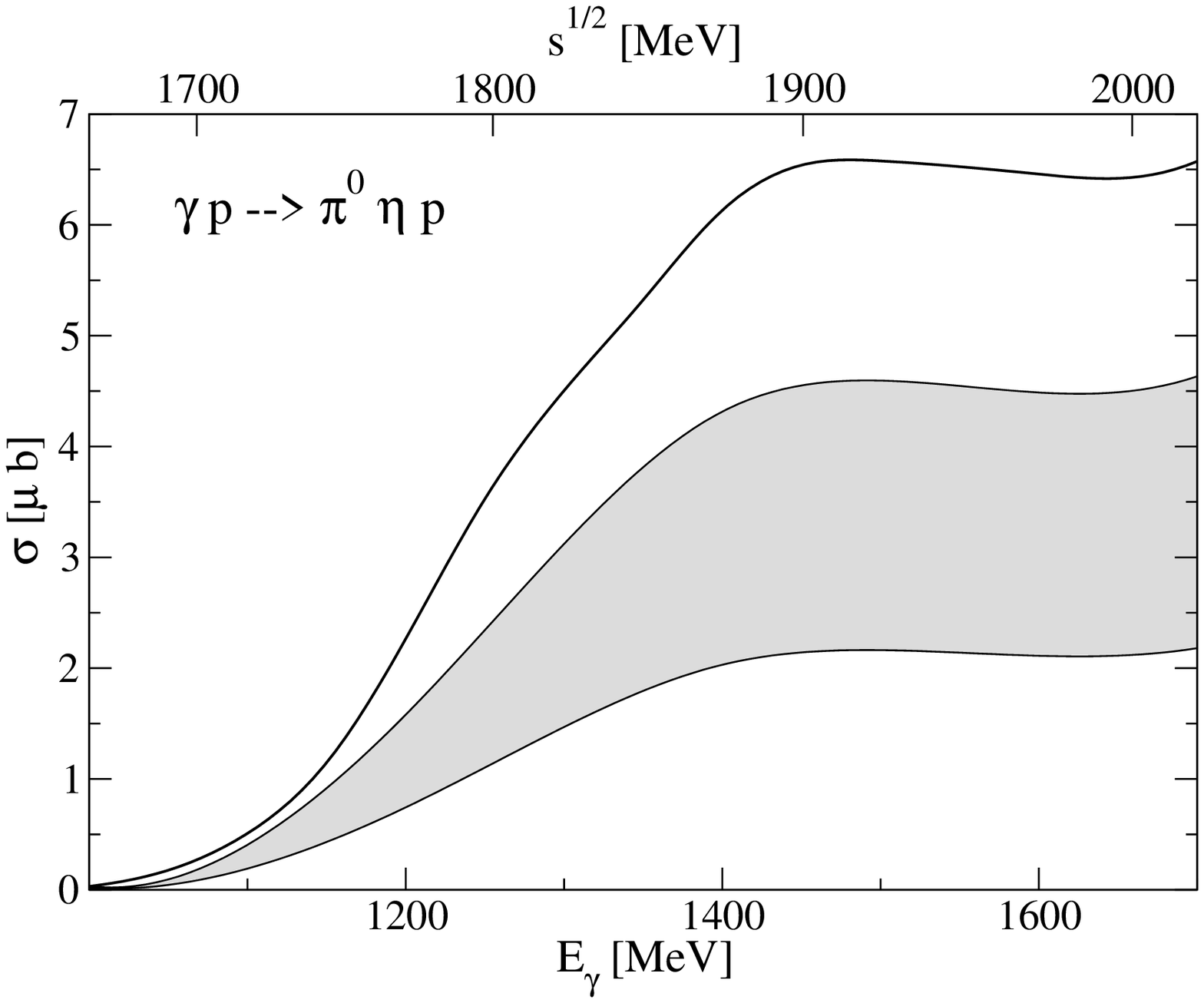}
\hspace*{0.2cm}
\includegraphics[width=4.3cm]{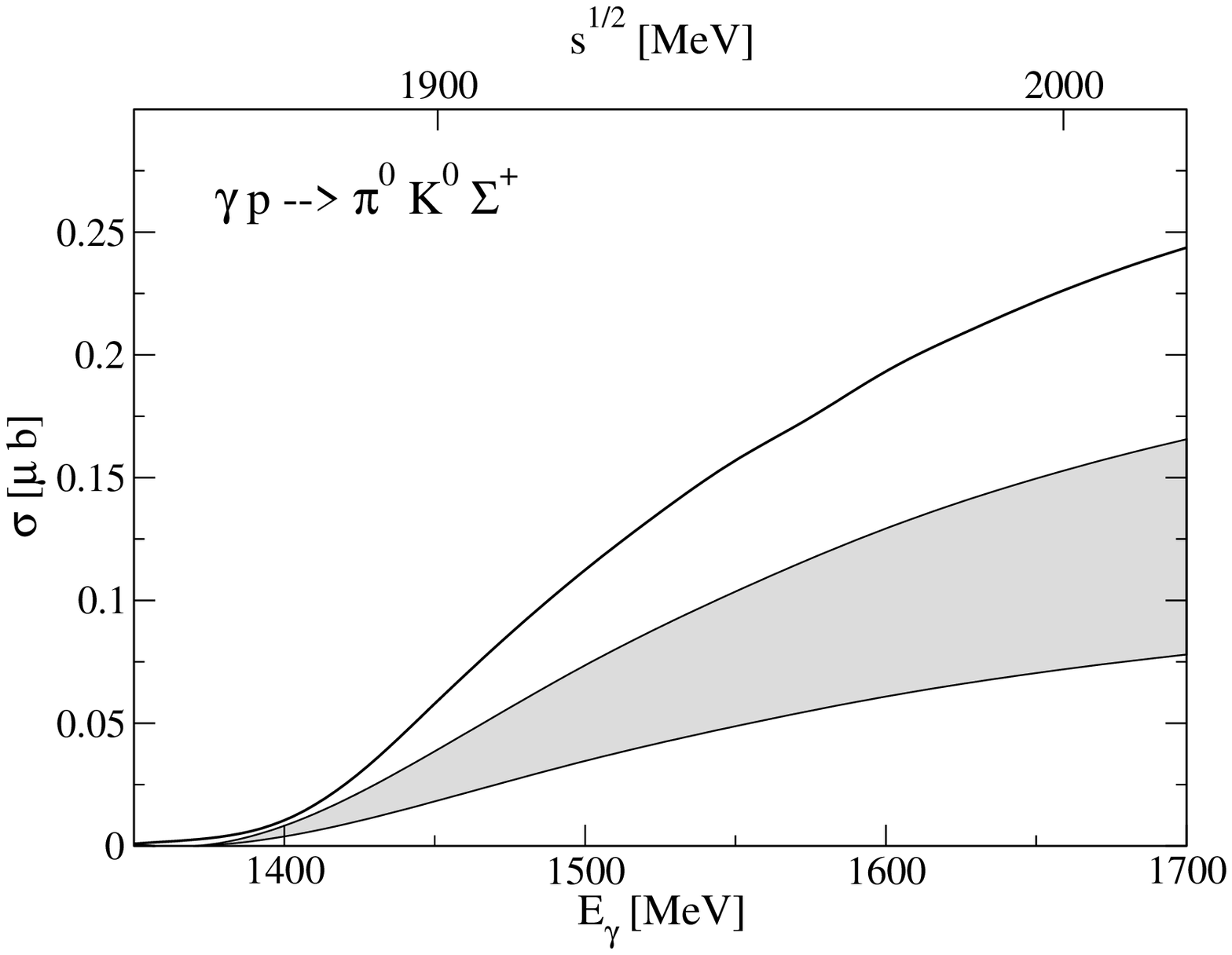}
\caption{Photoproduction of strange and $\eta$ particles. The gray bands are the
reuslts of \cite{delta}.}
\label{fig:photo}
\end{figure}  

\section{$\Lambda(1520)$ and $\Sigma(1385)$ in the nuclear medium}

\begin{figure}[h]
\begin{center}
\includegraphics[clip=true,width=0.5\columnwidth,angle=0.]
{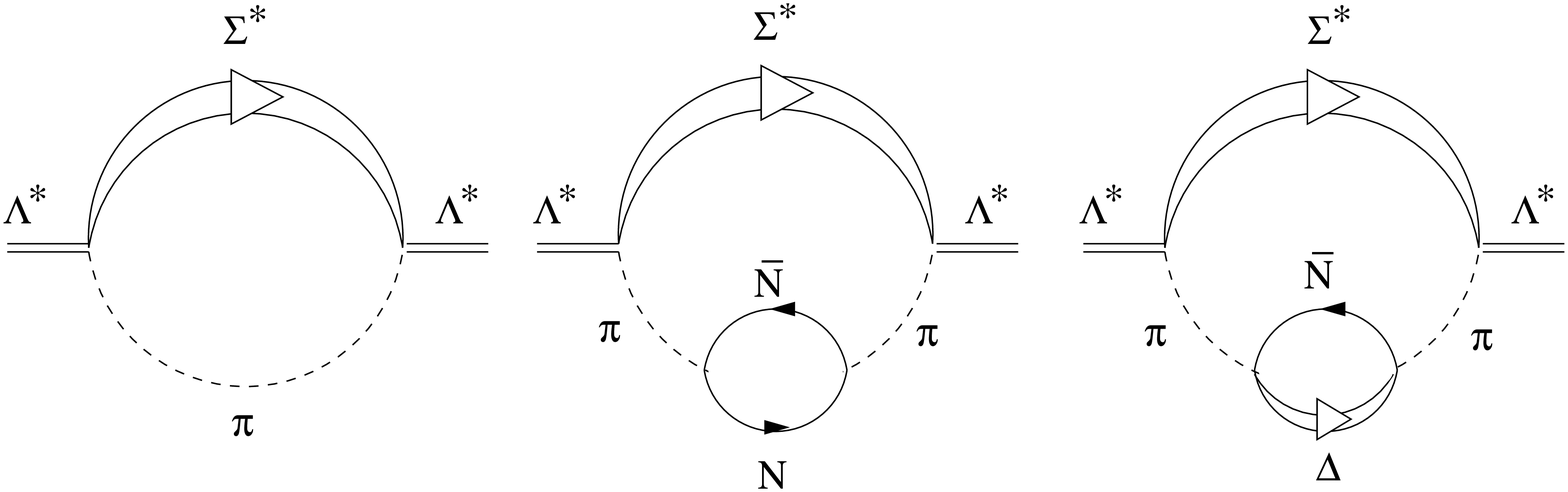}
\caption{ The renormalization of $\Lambda(1520)$
  in the nuclear medium.
}
\label{lambmed}
\end{center}
\end{figure}

The coupling of the $\Lambda(1520)$ to the $\pi \Sigma^*$ state is very large
but the decay is practically suppressed because there is no phase space for the
decay, up to a small overlap considering the width of the resonances.  However,
when we consider the decay in the medium, the pion can excite a $ph$ with energy
starting from zero, and then one gains 140 MeV of phase space for the decay,
see Fig.~ \ref{lambmed}.
This reminds one of the situation with the mesonic and nonmesonic decay of the
$\Lambda$ in a nucleus, in $\Lambda$ hypernuclei, where the mesonic decay is 
essentially forbidden by Pauli
blocking and then the $ph$ excitation of the pion gives rise to the nonmesonic
decay, which is dominant in the nucleus. The calculations have been done in 
\cite{murat} and the modifications to the $\Lambda(1520)$ width coming from the
$\pi \Sigma$ and $\bar{K} N$ in the nuclear medium are also taken into account.
In that work the selfenergy of the  $\Sigma(1385)$ is also evaluated and used to
calculate the selfenergy of the $\Lambda(1520)$.  One finds a substantial
increase in the width of the $\Sigma(1385)$, of more than a factor two for
nuclear matter density,  and a small shift of the mass. These results are consistent
with another independent and quite different evaluation of the  $\Sigma(1385)$
selfenergy done in \cite{laura}, where it appears as a byproduct of the study of
the p-wave $\bar{K} N$ interaction in the nuclear medium where the $\Sigma(1385)$
pole term is one of the important ingredients.

\begin{figure}[h]
\begin{center}
\includegraphics[clip=true,width=0.99\columnwidth,angle=0.]
{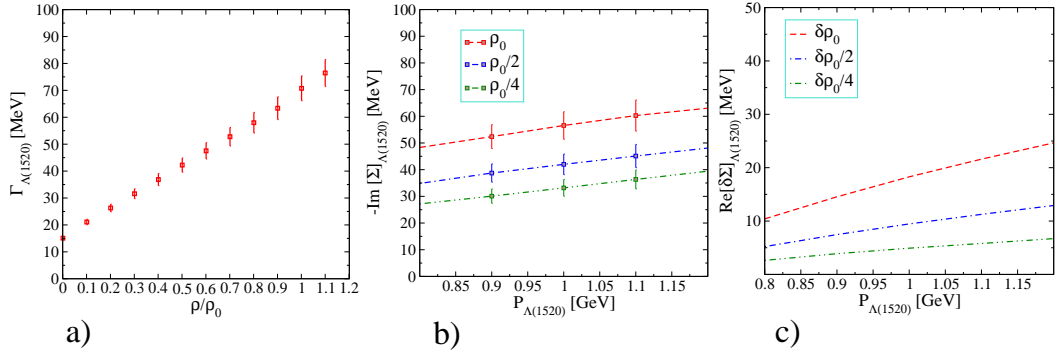}
\caption{ 
Values with theoretical uncertainties for the width of the $\Lambda(1520)$
at rest in the medium, including the free width, 
as function of the nuclear matter density
$\rho/\rho_0$ (a).
The 
imaginary (b) and the vacuum subtracted real (c) parts of the
$\Lambda(1520)$ selfenergy as a function of the $\Lambda(1520)$ three momentum.
}
\label{lambmedres}
\end{center}
\end{figure}

In fig. ~\ref{lambmedres} we show results for the $\Lambda(1520)$ width in medium.
The figure to the left shows the width as a function of the nuclear density. We
can see that at $\rho= \rho_0$ the width is about five times bigger than the free
width.  This is is a spectacular increase which should not go unnoticed in
experiments and some suggestions have already been done in \cite{muratluis}.

\section{$\omega$ in the nuclear medium}
\begin{figure}[h]
\begin{center}
\includegraphics[clip=true,width=0.5\columnwidth,angle=-90.]
{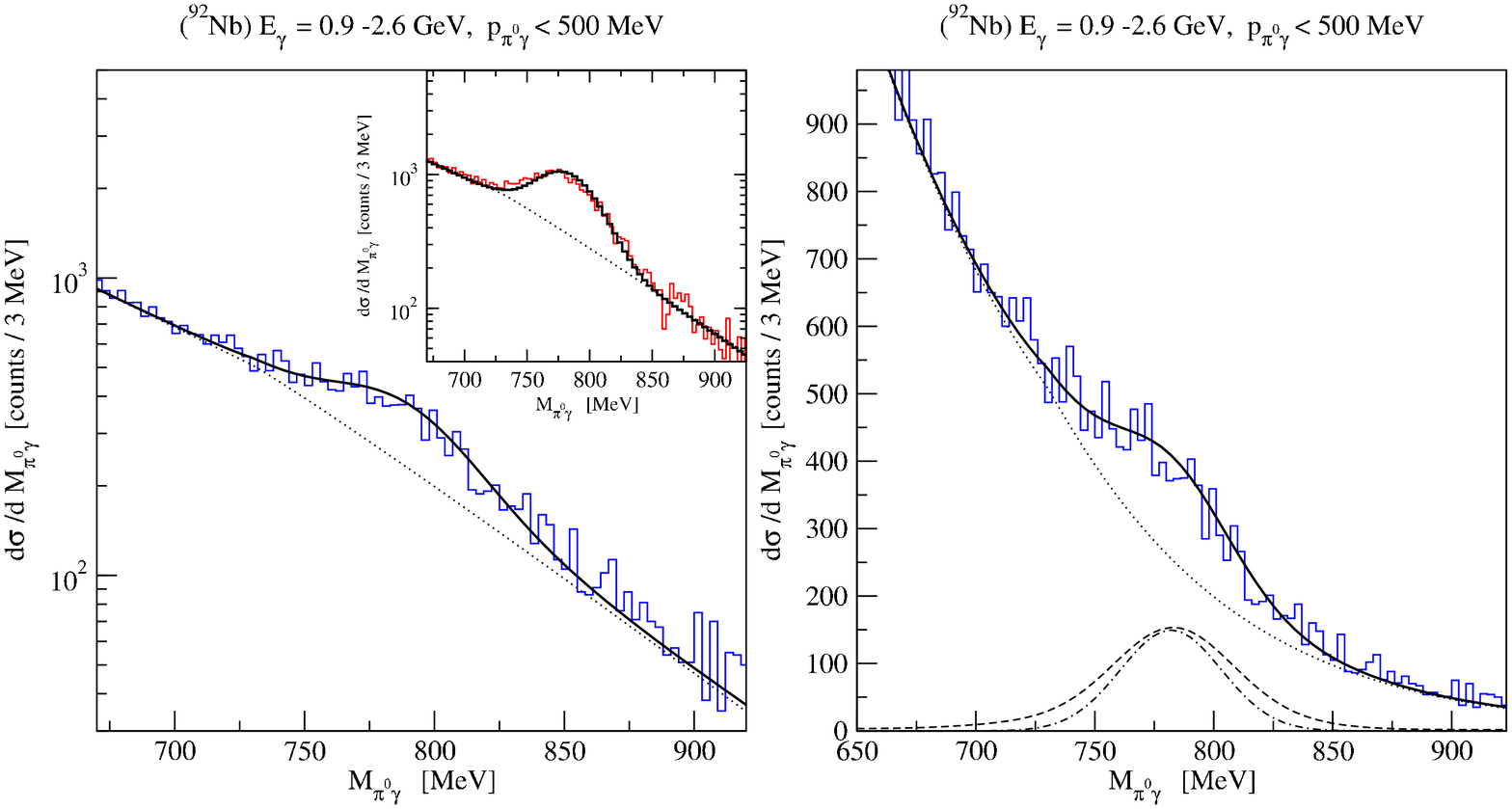}
\caption{Background (dotted line) and $\omega$ signal in the proton (inset) and the nucleus.
}
\label{omega}
\end{center}
\end{figure}

This issue is an interesting one and related to the general topic of the possible
mass shift of the vector masses in the nuclear medium predicted by some theories. 
 An interesting experiment was done in \cite{trnka} by means of the 
 $\gamma A \to \omega (\pi^0 \gamma) A'$ reaction, where the $\omega$ was
 identified from the invariant mass of the $\pi^0 \gamma$ from its decay into
 this channel.  A shift of the invariant mass  with respect to the one in the
 $\gamma p$ reaction was attributed in \cite{trnka} to a shift of the $\omega$
 mass in the medium. In Ref.  \cite{eli} a reanalysis of the process is done by
 means of a Monte Carlo simulation procedure that traces the interaction of all the
 particles inside the nucleus and considers a possible shift of mass and width of the
 $\omega$ in the nucleus which is fitted to the data. An important novelty of this 
 work is that it chooses
 the background in the nucleus proportional to the one in the proton reaction in
 the energy region around the $\omega$ mass, while in \cite{trnka} the background
 is manifestly changed, and increased at higher masses, in such a way that 
 there are no events in the region of high invariant masses. Hence, the shift to
 lower $\omega$ masses claimed in \cite{trnka} can be directly linked to the choice 
 of background. On the contrary, in \cite{eli} we find that the data demand a
 large $\omega$ width, of the order of $90$  MeV at normal nuclear matter
 density, but there is no need for a shift of the mass. In Fig.~ \ref{omega}   
 we show the results obtained in \cite{eli} together with the backgrounds in the
 nucleus of $^{92}Nb$ and on the proton (inset).  In addition, in \cite{hideko}
 it is found that should the $\omega$ potential be sufficiently attractive to
 support bound $\omega$ states in nuclei, one needs a better resolution than the
 one available at ELSA in  \cite{trnka} to see the states in the $\gamma A$ 
 reactions looking for p in coincidence at small angles.  These findings, and
 others discussed in \cite{hideko} are very useful to help interpret properly
 the results obtained in experiments.

\section{Conclusions}
We have presented a selection of recent results on hadron physics and hadrons in
the nuclear medium which, once again, show the adequacy of chiral unitary theory to
address the world of hadron physics at intermediate energies and subsequent
problems of hadrons in a nuclear environment. We could see how some properties
of resonances
like  the $\Lambda(1520)$  and $\Delta(1700)$ can be naturally
traced to the interpretation of these resonances as dynamically generated from
the interaction of the octet of pseudoscalars with the decuplet of baryons. We
also showed some reactions supporting the findings of chiral theory about the
existence of two $\Lambda(1405)$ states.  We then showed how this nature has
immediate repercussions for the behaviour of these resonances in nuclei, as a
consequence of which the $\Lambda(1520)$ gets a very large width in the medium,
of the order of five times bigger than its free width at normal nuclear matter
density.  Finally we also presented some results of a reanalysis of the results
of $\omega$ production in nuclei and found that, contrary to what was assumed
from  a previous analysis, a recent work subtracting a more appropriate
background, and imposing the constraints of a large width in the medium obtained
in the same experiment,  showed that the results can be interpreted in terms
of this enlarged width without invoking any shift in the mass.

\section*{Acknowledgments}
One of the authors (E. O.) thanks the Yukawa Institute for Theoretical 
Physics at Kyoto University, where some of the ideas exposed here were matured
 during the YKIS2006 on "New Frontiers on QCD". 
This work is partly supported by DGICYT contract
number FIS2006-03438, the Generalitat Valenciana. This research is  part of the EU Integrated
Infrastructure Initiative  Hadron Physics Project under  contract number
RII3-CT-2004-506078.

\end{document}